\documentclass[10pt]{article}
    \usepackage[english]{babel}
    \usepackage[margin=1in]{geometry}
    \usepackage{background}
\usepackage{lineno}

\usepackage[morefloats=7]{morefloats}
\usepackage[compact]{titlesec}
\usepackage{graphicx}
\usepackage{amssymb}
\usepackage{float}
\usepackage{algorithmic}
\usepackage{listings} 
\usepackage{framed}
\usepackage{amsmath}
\usepackage[section]{placeins}
\usepackage[justification=centering]{caption}
\usepackage[sort&compress, numbers]{natbib} 
\usepackage{hyperref}
\usepackage{breakurl}

\usepackage[linesnumbered,ruled,vlined]{algorithm2e}
\usepackage{subcaption}
\usepackage{makecell}
\usepackage{array}
\newcolumntype{L}{>{\centering\arraybackslash}m{3cm}}

    \SetBgScale{4}
    \SetBgColor{gray}
    \SetBgAngle{90}
    \SetBgPosition{-1.5,-10}

 \usepackage{fancyvrb}
\usepackage{authblk}

\title{GPU-Based Fuzzy C-Means Clustering Algorithm for Image Segmentation}
\author[1]{Mishal Almazrooie}
\author[1]{Mogana Vadiveloo}
\author[1,2]{Rosni Abdullah\thanks{thanks}}
\affil[1]{School of Computer Sciences, Universiti Sains Malaysia, 11800 USM, Pulau Pinang, Malaysia}
\affil[2]{National Advanced IPv6 Center (NaV6), Universiti Sains Malaysia, 11800 USM, Pulau Pinang, Malaysia}

\begin{document}
%\twocolumn[
%\begin{@twocolumnfalse}
\maketitle

 %   {\centering

 %   {\bfseries\Large GPU-Based Fuzzy C-Means Clustering Algorithm for Image Segmentation\bigskip}

 %   Mishal Almazrooie\textsuperscript{1} , Mogana Vadiveloo\textsuperscript{1} , and Rosni Abdullah\textsuperscript{1,2}\\
 %     {\itshape
 %   \textsuperscript{1}School of Computer Sciences, Universiti Sains Malaysia, 11800 USM, Pulau Pinang, Malaysia, \\
  %  \textsuperscript{2}National Advanced IPv6 Center (NaV6), Universiti Sains Malaysia, 11800 USM, Pulau Pinang, Malaysia\\
  %  \normalfont (Dated: January 01, 2016)
    %  }
 %   }

    \begin{abstract}
In this paper, a fast and practical GPU-based implementation of Fuzzy C-Means(FCM) clustering algorithm for image segmentation is proposed. First, an extensive analysis is conducted to study the dependency among the image pixels in the algorithm for parallelization. The proposed GPU-based FCM has been tested on digital brain simulated dataset to segment white matter(WM), gray matter(GM) and cerebrospinal fluid (CSF) soft tissue regions. The execution time of the sequential FCM is 519 seconds for an image dataset with the size of 1MB. While the proposed GPU-based FCM requires only 2.33 seconds for the similar size of image dataset. An estimated 245-fold speedup is measured for the data size of 40 KB on a CUDA device that has 448 processors.
    \bigskip

    \noindent Image segmentation, Fuzzy C-Means, Parallel algorithms, Graphic Processing Units (GPUs), CUDA 

    \end{abstract}
%\end{@twocolumnfalse}
%]
    % the article contents goes here
    %\blinddocument % just to produce some filler text
\FloatBarrier
\section{Introduction}
\label{intro}
Image segmentation has been one of the fundamental research areas in image processing. It is a process of partitioning a given image into desired regions according to the chosen image feature information such as intensity or texture. The segmentation is used with application in the field of medical imaging, tumors locating and diagnosis. Over the past few decades, as image segmentation has gained much interest, various segmentation techniques have been proposed, each of which uses different induction principle.\\   
\indent Clustering is one of the most popular techniques used in image segmentation. In clustering, the goal is to produce coherent clusters of pixels \cite{Chattopadhyay}. The pixels in a cluster are as similar as possible with respect to the selected image feature information. While the pixels belong in the adjacent clusters are significantly different with respect to the same selected image feature information \cite{Chattopadhyay}. There are variants of clustering algorithms have been used widely in image segmentation and they are K-Means \cite{Tou}, Fuzzy C-Means (FCM) \cite{Bezdek}, and ISODATA \cite{Ball}. \\
\indent In the last decades, FCM has been very popularly used to solve the image segmentation problems \cite{Shen}; \cite{Vadiveloo}. It is a fuzzy clustering method that allows a single pixel to belong to two or more clusters. The introduction of fuzziness makes this algorithm to able to retain more information from the original image than the crisp or hard clustering algorithms \cite{Shen}; \cite{Vadiveloo}. However this sequential FCM becomes computationally intensive when segmenting large image datasets \cite{Vadiveloo}. In such a case, the algorithm becomes very inefficient.\\   
\indent One-way to improve the performance of the FCM clustering algorithm is to use parallel computing methods. Initially, Graphic Processing Units (GPUs) were specific-purpose processors that only manipulate and accelerate the creation of images intended for output to a display. However, GPUs have recently shifted to general-purpose processors (GPGPUs) to solve general concerns, such as scientific and engineering problems. Data parallelism on a GPU is a powerful parallel model. In this paper, a fast and practical parallel FCM approach on GPGPU is presented and discussed. \\
\indent This paper is organized as follows: Section 2 provides a background of FCM algorithm and the parallel technology used. Section 3 presents related works. The proposed method is explained in detail in Section 4. The experimental results are presented and discussed in Section 5. Finally, Section 6 provides the conclusion and suggestions for future works.
\FloatBarrier
\section{Preliminaries}
\label{sec:Preliminaries}
In the first sub-section, a brief introduction on Fuzzy C-Means (FCM) algorithm is presented. While in the following sub-section the parallel technology used in this work namely on General Purpose Computing on Graphics Processing Units (GPGPU) data parallelism is discussed. 
\FloatBarrier
\subsection{Fuzzy C-Means Algorithm}
\label{sec:fcm}
Fuzzy C-Means was developed by \cite{Bezdek}. It is an iterative optimization that minimizes the objective function defined in \ref{eq1111}. The objective function consists of two main components $u$ and $v$.~$u_{ij}$ is the membership function of a pixel, $x_i$. It represents the probability that $x_i$ may belong to a cluster. The $u_{ij}$ is dependent on the distance function, $d_{ij}$. $d_{ij}$ is the Euclidean distance measure between the pixel $x_i$ and each cluster center,  $v_j, d_{ij}=|| x_i - v_j ||$. $m$ is a constant that represents the fuzziness value of the resulting clusters that are to be formed; $1\leq m\leq \infty$.
\begin{equation}\label{eq1111}
\large J_{i}= \sum\limits_{i=1}^N \sum\limits_{j=1}^c u_{ij}^m {||x_i - v_j||}^2 \\ 
\end{equation}
\noindent with respect to:
\begin{equation}\label{eq112}
\begin{split}
\large \sum\limits_{j=1}^c  u_{ij} = 1, 1\leq i\leq n \\ 
0 < \sum\limits_{i=1}^n  u_{ij} < n, 1\leq j\leq c \\
\sum\limits_{i=1}^c \sum\limits_{i=1}^n u_{ij} = n .
\end{split}
\end{equation}
\indent In image clustering, the most commonly used feature is the grey level or intensity value of the image being segmented. Therefore, the objective function,$ j_m$ in \ref{eq1111} is minimized when higher membership value is assigned to pixels with intensity values close to a cluster center of the corresponding cluster, while lower membership value is assigned to pixels whose intensities are far from the cluster center.  
\begin{equation}\label{eq113}
\large v_{i}= \frac{\sum\limits_{i=1}^N u_{ij}^m. x_{i}}{\sum\limits_{i=1}^N u_{ij}^m} 
\end{equation}
\begin{equation}\label{eq114}
\large u_{ij}=\frac{1}{\sum\limits_{k=1}^c \bigg[{\frac{||x_{i}-v_{j}||}{||x_{i}-v_{k}||}\bigg]^\frac{2}{m-1}}}
\end{equation}
\indent Starting with random initialization of the membership values for each pixel from the manually selected clusters, the clusters are converged by recursively updating the cluster centers and membership function in \ref{eq113} and \ref{eq114}. This is to minimize the objective function in \ref{eq1111}. Convergence stops when the overall difference in the membership function between the current and previous iteration is smaller than a given epsilon value, $\varepsilon$. After the convergence, deffuzzifaction is applied. Each pixel is assigned to a specific cluster according to the maximal value of its membership function. The steps of the Fuzzy C-Means algorithm are illustrated in Algorithm \ref{algo1C}. 

{\LinesNumberedHidden
\begin{algorithm}
\DontPrintSemicolon % Some LaTeX compilers require you to use \dontprintsemicolon instead
\textbf{Assumptions: Image is transformed into feature space.}\\

Step 1: Initialize the number of clusters $c$,  $m =2$, and $\varepsilon = 0.005$\\
Step 2: Initialize the membership function, $u_{ij}$  randomly.\\
Step 3: \Repeat {$||u^{k+1}_{ij}-u^{k}_{ij}|| < \varepsilon $ } {
 Update the cluster center, $v_i$ using Equation \ref{eq113}\\
 Update the membership function $u_{ij}$ using Equation \ref{eq114}\\
  }
\caption{{\sc Fuzzy C-Means algorithm} }
\label{algo1C}
\end{algorithm}}
\FloatBarrier
\subsection{Data Parallelism on GPGPU}
\label{sec:gpu}
\begin{figure}[H]
\begin{center}
\includegraphics[scale=0.55]{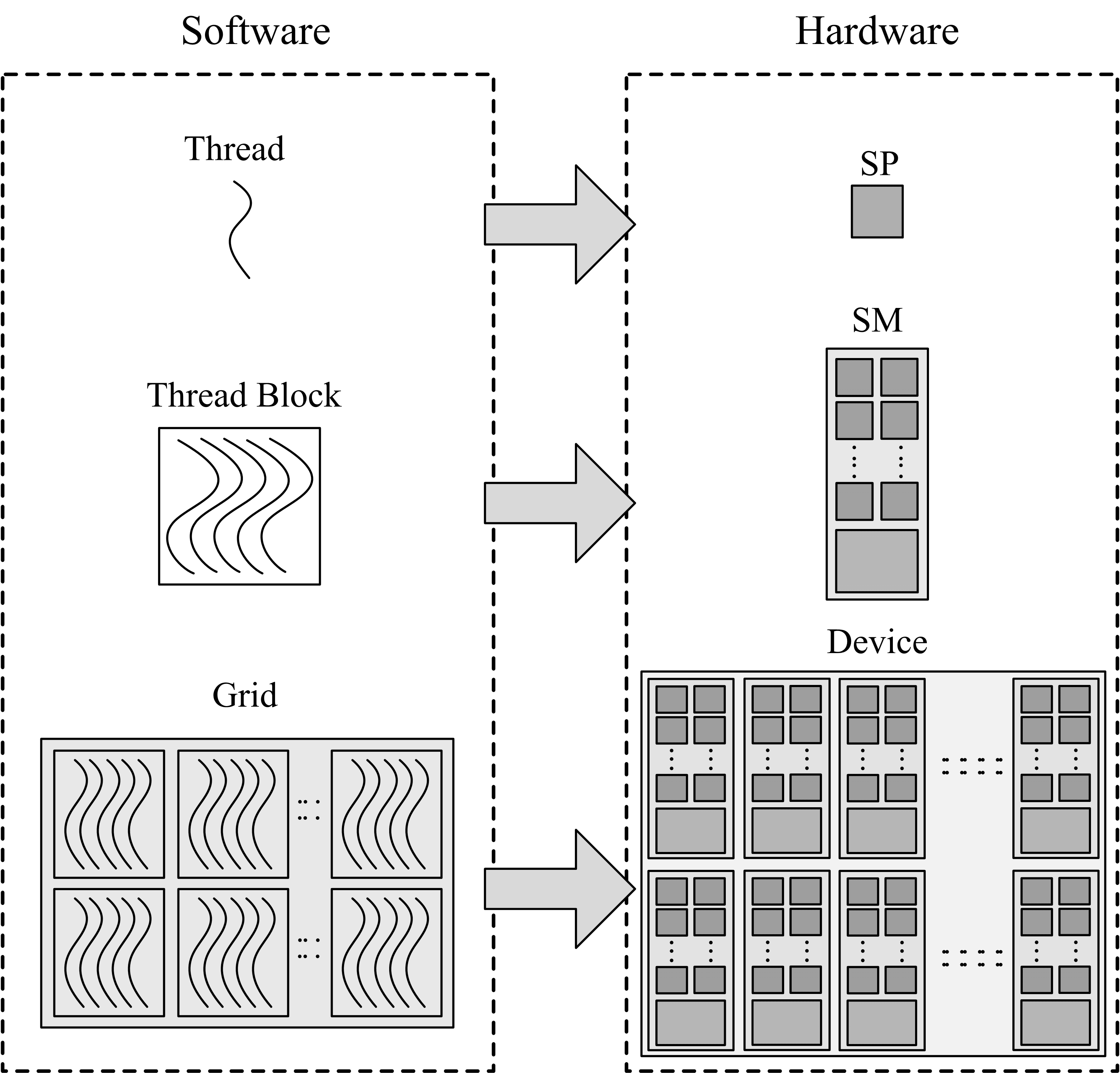}
\caption{CUDA Execution Model on GPGPU}
\label{executionmodel}
\end{center}
\end{figure}

Initially, GPU was a hardware equipped with a processor specifically designed to accelerate graphic processing. Eventually, GPU applications were extended to general-purpose computations. At present, GPGPU is used in many applications typically performed using a CPU, such as analytic, engineering, and scientific applications \cite{Farber}. With the release of the massively parallel architecture called CUDA in 2007 from NVIDIA, GPUs have become widely accessible \cite{Kirk2010}.\\
\indent A GPU is a processor or a multiprocessor device that has hundreds or even thousands of cores called scalar processors (SPs), which are arranged in groups named streaming multiprocessors (SMs), as shown in the left side of Fig. \ref{executionmodel}. Moreover, GPUs have different kinds of memories: global, local, texture, constant, shared, and register memories. Global, constant, and texture memories are accessible to all threads in the grid. Shared memory is visible to threads within one CUDA block. It is faster than the global memory but is limited by size. Register memory is visible to the thread that initialized the said memory and lasts for the lifetime of that thread.\\
\indent CUDA is the parallel programming model used for NVIDIA GPGPUs. CUDA can increase the performance by harnessing the power of a GPU device. Thousands of threads can be executed concurrently using CUDA on GPGPU. The execution model of CUDA on NVIDIA devices is shown in Fig. \ref{executionmodel}.
%\end{minipage}
\section{Related works}
\label{sec:Related works}
Li et al. proposed an Fuzzy C-Means (FCM) algorithm based on GPU \cite{Li2014}. They modified the sequential FCM algorithm, such that the calculations of the membership and cluster center matrices are not comparable to the sequential one. They have FCM on GPU using CUDA. The empirical results obtained by Li et al. showed that the proposed parallel FCM on GPU is more efficient than the sequential FCM. Instead of efficiency, they claimed that the proposed method exhibits improvement in the quality of the GPU segmented image. The authors achieved a 10-fold speedup with the proposed parallel FCM on NVIDIA GTX 260 device compared with the sequential FCM for natural images sized from 53kb to 101kb.\\
\indent Mahmoud et al. presented a GPU-based brFCM for medical images segmentation \cite{Mahmoud2015}. The brFCM is a faster variant of the sequential FCM \cite{Eschrich2003}. The GPU-based brFCM is implemented on different GPGPU cards. Mahmoud et al  showed that the GPU-based brFCM  has a significant improvement over the parallel FCM in \cite{2014Soroosh}. The achieved speedup is up tp 23.42 fold faster than parallel FCM in \cite{2014Soroosh} for medical images of 350x350 and 512x512 dimensions.\\
\indent Shalom et al. proposed a scalable FCM based on graphic hardware \cite{Shalom2008}. On two different graphic cards, the results show that the proposed GPU-based FCM algorithm is more efficient and faster than the sequential FCM. The authors succeeded in reaching a 73-fold speedup on NVIDIA GeForce 8500 GT. Amazingly, a 140-fold speedup was achieved on NVIDIA GeForce 8800 GTX compared with sequential FCM for 65k yeast gene expression data set of 79 dimension.\\
\indent Rowinska and Goclawski proposed a CUDA-based FCM algorithm to accelerate image-segmentation \cite{Rowi2012}. The proposed method has been tested on polyurethane foam with fungus color images and was compared with the sequential FCM implemented using C++ and MATLAB. The authors achieved a 10-fold speedup of their parallel proposal compared with the FCM implemented in C++ for object area of 310k pixels, and a 50- to 100-fold speedup compared with the FCM implemented in MATLAB for object area of 260k pixels. A comparison of our work and the previous related works is summarized in Table \ref{relatedworks}. 
\begin{table*}[t]
\centering
\caption{Comparison of our work and previous related works} % title of Table
\begin{tabular}{c|L|L|c}
\textbf{Work by}& \textbf{Method} & \textbf{Image dataset} & \textbf{Speedup}\\\hline % inserts table 
%heading
$ \begin{matrix} \text{Li et al.} \\ \text{\cite{Li2014}}\end{matrix}$ & \multicolumn{1}{m{3cm}|}{Modified the original FCM algorithm and then parallelized it on GPGPU}&  \multicolumn{1}{m{3cm}|}{Natural images (from 53kB to 101kB)} & 10x  \\\hline
$ \begin{matrix} \text{Mahmoud et al.} \\ \text{\cite{Mahmoud2015}}\end{matrix}$ & \multicolumn{1}{m{3cm}|}{Parallelized brFCM the variant of FCM algorithm on GPGPU}&  \multicolumn{1}{m{3cm}|}{Medical images (Lung CT with the dimesion of 512x512; Knee MRI with the dimension of 350x350)} & \makecell{23x faster\\ than in \cite{2014Soroosh} } \\\hline
$ \begin{matrix} \text{Shalom et al } \\ \text{\cite{Shalom2008}}\end{matrix}$ & \multicolumn{1}{m{3cm}|}{Proposed a scalable FCM GPU-based implementation}&  \multicolumn{1}{m{3cm}|}{Yeast gene expression data set (79 dimension with 65K genes)} & 140x  \\\hline
$ \begin{matrix} \text{Rowinska et al.} \\ \text{\cite{Rowi2012}}\end{matrix}$ & \multicolumn{1}{m{3cm}|}{Presented a CUDA-based FCM algorithm to accelerate image segmentation}&  \multicolumn{1}{m{3cm}|}{Polyurethane foam with fungus color images (object area of 310k pixels)} & 10x  \\\hline
This paper & \multicolumn{1}{m{3cm}|}{A parallel FCM approach on GPGPU using CUDA  }&  \multicolumn{1}{m{3cm}|}{Digital brain phantom simulated dataset (from 20kB to 1000kB)} & $ \begin{matrix} \text{speedup} \\ \text{up to 245x} \end{matrix}$    
\end{tabular}
\label{relatedworks} % is used to refer this table in the text
\end{table*}
\FloatBarrier
\section{The Proposed Method}
\label{sec:FCM Parallel Approach}
\begin{figure*}[!t]
\begin{center}
\includegraphics[scale=0.71]{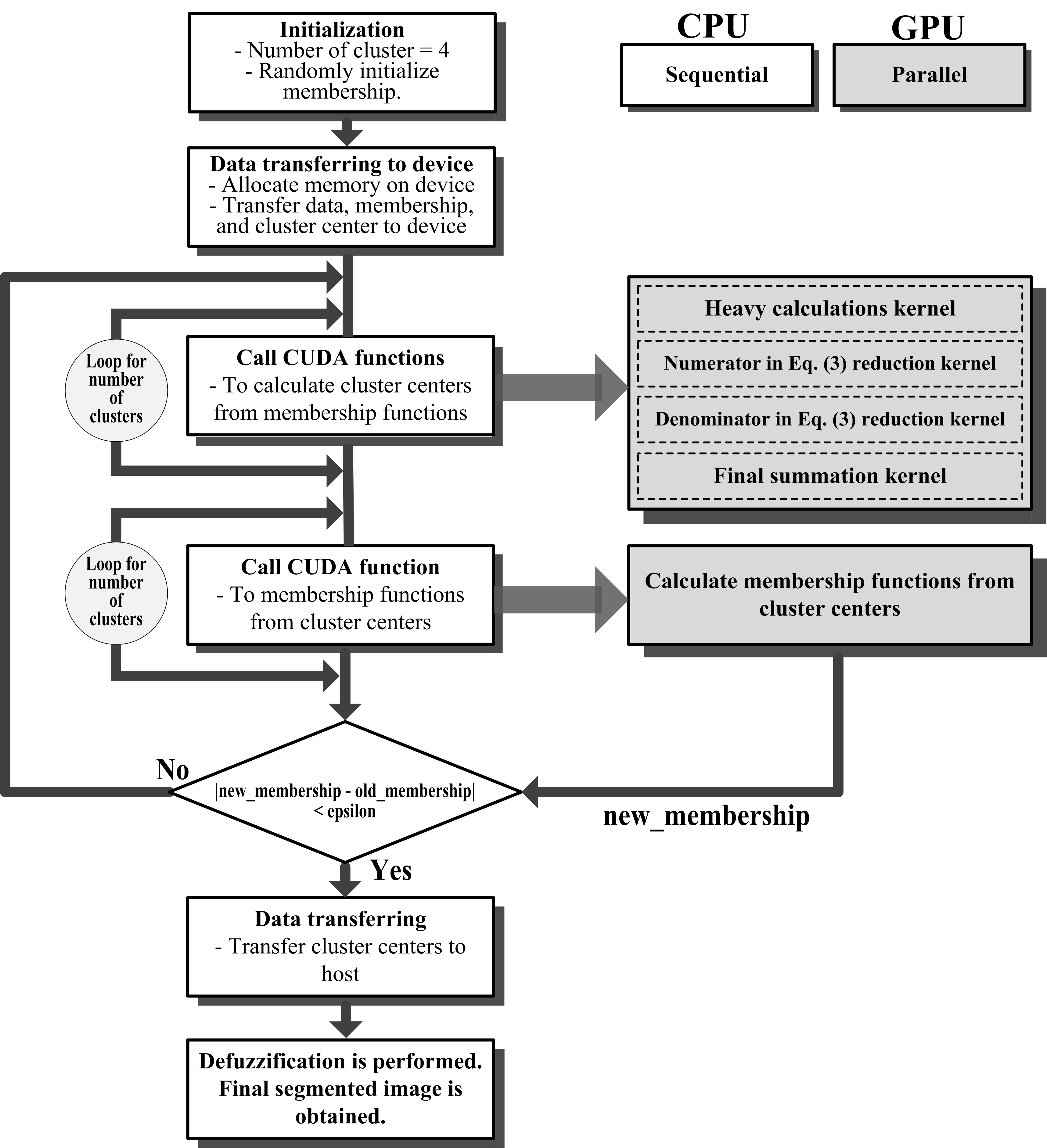}
\caption{The Block Diagram of The Proposed Parallel Fuzzy C-Means}
\label{block}
\end{center}
\end{figure*}

The sequential FCM algorithm has been subjected to extensive analysis in order to find out where the algorithm exhibits parallelism that we might exploit in the parallel design. The strongest data dependency in the FCM algorithm is the steps where the total summation calculation is required, as illustrated in step 3 in the sequential FCM (Algorithm \ref{algo1C}). For instance, two sigma operations are needed to calculate the cluster centers as shown in Equation \ref{eq113}. Such a strong dependency makes parallelizing the sequential algorithm infeasible. According to Bernstein’s conditions \cite{Bernstein}, this type of dependency is called output dependence. In parallel computing, the reduction method is an efficient approach to remove output dependence.\\
\indent The proposed parallel FCM design consists of two main parts: a sequential part executed on the CPU (host) and a parallel part executed on the GPU (device). Fig. \ref{block} shows the block diagram of the proposed work. The following sub-sections discuss each stages of the block diagram in Fig. \ref{block}. 
\FloatBarrier
\subsection{Initialization and data transferring}
\label{sec:Initialization}
As shown in Fig. \ref{block}, the first two steps are executed sequentially on the host. The membership is randomly initialized. The memories are allocated on the device global memory for the pixels of the image data, membership, and cluster centers. All the arrays are defined in a 1-D pattern.\\
\indent After defining memories on the device, all the data are transferred from host to device, and then the main program loop is started. Subsequently, the parallel kernels are called concurrently to manipulate the image pixels on the device.
\FloatBarrier
\subsection{Calculating cluster centers from membership functions}
\label{sec:Calculating cluster centers}
\begin{figure}[H]
\begin{center}
\includegraphics[scale=0.45]{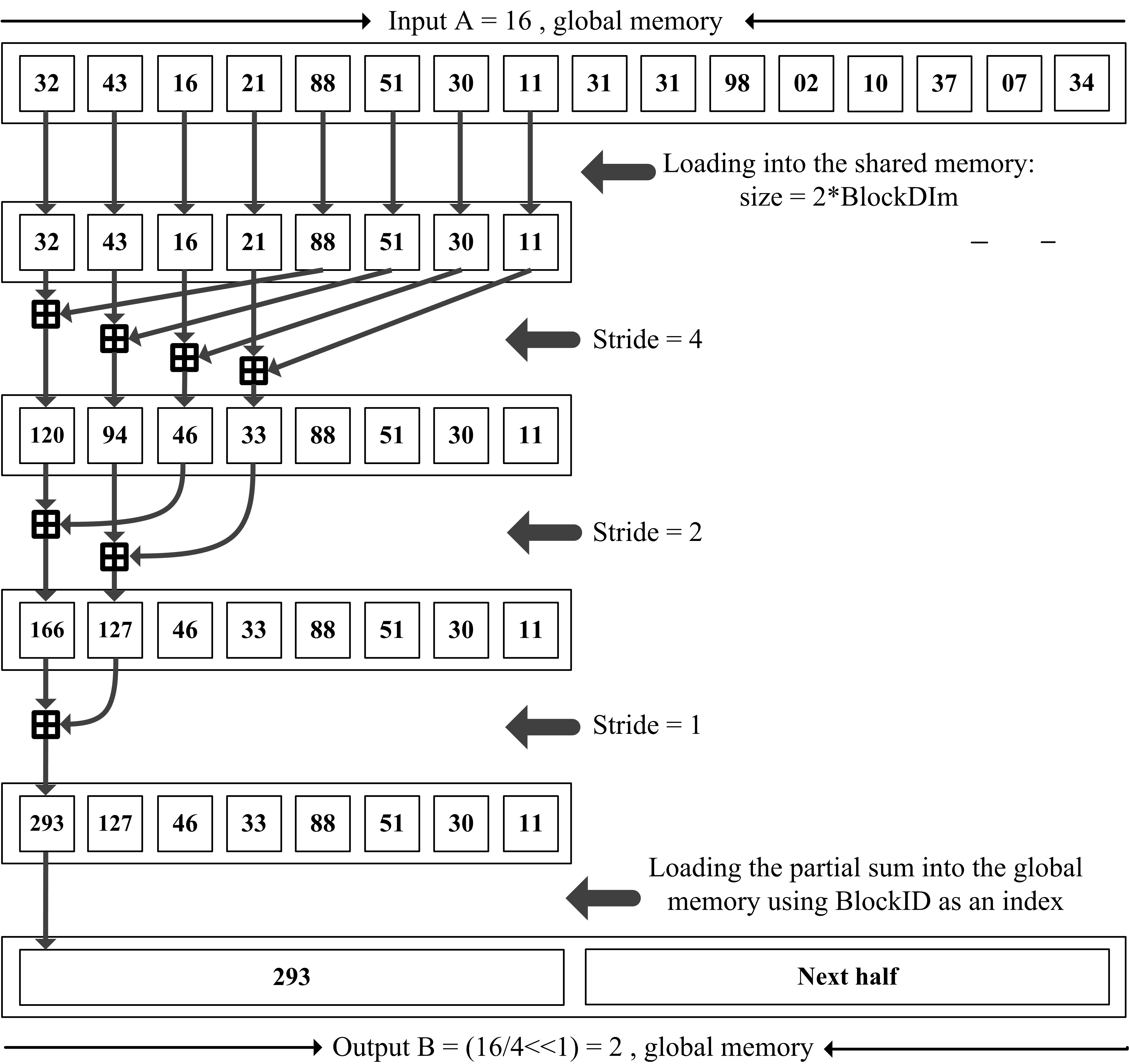}
\caption{Sum Reduction Example on GPGPU. There are four CUDA block dimension in this example.}
\label{reduction}
\end{center}
\end{figure}

The host calls four CUDA kernels one after another to calculate the cluster centers from memberships. The first CUDA kernel concurrently handles the heavy calculations, such as exponential, division, and multiplication of floating points for every pixel. At this step, the final summation is not included. The numerator and denominator of Equation \ref{eq113} are calculated separately for every pixel, and the results are stored in two different arrays in the device global memory. The number of spawned CUDA threads in this kernel is defined to be equal to the number of image pixels, such that every thread will handle one pixel.      
\indent The second CUDA kernel at this phase is the reduction kernel, which computes the partial summation of the numerator of Equation \ref{eq113}. The reduction technique is an efficient method to break down the dependency among the data. The computation complexity of the sequential addition of $n$ elements is $O(n)$. However, using parallel computing can significantly improve the computation complexity to $O(log~n)$ \cite{2014r3}\cite{2014r6}. Several CUDA reduction methods are available, such as in \cite{2014r1}\cite{2014r2}\cite{2014r3}. The CUDA reduction method used in this work is similar to \cite{2014r3} and is shown in Algorithm \ref{algo1}. First, a segment of the input is loaded into the device-shared memory. This device shared memory can facilitate fast access to the image pixels \cite{2014r4}\cite{2014r5}. The reduction process is then performed over the shared memory. Each calculated partial sum of every segment stored in the shared memory is loaded to the output in the global memory. As illustrated in Algorithm \ref{algo1}, the CUDA block ID (blockIdx.x) is used as an index to store the partial sum from the device-shared memory to the global memory. Fig. \ref{reduction} demonstrates the reduction process performed on GPGPU using shared memory.
\begin{algorithm*}
\DontPrintSemicolon % Some LaTeX compilers require you to use \dontprintsemicolon instead
\KwIn{A large set $A=\{a_1, a_2, \ldots, a_n\}$ where $n = pixels$}
\KwOut{A reduced small set $B=\{b_1, b_2, \ldots, b_m\}$ where $m = n / blockDim<<1$}
$global\_idx \gets blockIdx.x*blockDim.x + threadIdx.x$\;
$local\_idx \gets threadIdx.x$\; 
$start \gets  2*blockIdx.x*blockDim.x$\;
\_\_shared\_\_  $partialSum[2*MAX\_THREAD]$\;
\textbf{//Loading segment from the input into the shared memory:}\;
\If{$(start + local\_idx) < n$}
    {
        $partialSum[local\_idx] = A[start + local\_idx]$\;      
    }
\Else
    {       
        $partialSum[local\_idx] = 0.0$\;
    }
\If{$(start + local\_idx+ blockDim.x) < n$}
    {
        $partialSum[local\_idx + blockDim.x] = A[start + local\_idx + blockDim.x]$\;      
    }
\Else
    {       
        $partialSum[local\_idx + blockDim.x] = 0.0$\;
    }
\textbf{//Reduction over the device shared memory:}\;
\For{$stride \gets blockDim.x$ \textbf{to} $0$ ; $stride /= 2$} {
  \If{$local\_idx < stride$} {
    $partialSum[local\_idx] += partialSum[local\_idx + stride]$\;
  }
}
\textbf{//Storing the output into the device global memory:}\;
\If{$local\_idx == 0 \&\& (global\_idx*2) < n$} {
    $B[blockIdx.x] = partialSum[local\_idx]$\;
}
\caption{{\sc Sum Reduction on GPGPU Using CUDA} }
\label{algo1}
\end{algorithm*}
The actual reduction for the illustrated example in Fig. \ref{reduction}, reduces the addition operations from adding 16 elements to only 2 elements. Another example from the conducted experiments of this work is an image with a size of 1 MB (1048576 bytes) that was reduced to $(1048576/128<<1)$, which equals 4 KB (4096 bytes).\\
\indent The third kernel to be called in this phase (calculating cluster centers from the membership function phase) is another reduction kernel that calculates the partial sum of the denominator of Equation \ref{eq113}. Finally, the last CUDA kernel calculates both final summations from the previous two kernels and computes the final result. Only one thread is defined for this kernel. The reason for this one thread kernel is that instead of transferring the reduced arrays from the previous kernels to the host memory to calculate the final summations, in this proposed method the device is allowed to carry out the final summation only with one thread. Lastly, all the previous four CUDA kernels are called in iterative loops that are equal to the predefined number of clusters. This is to calculate the cluster centers from the membership functions as shown in Fig. \ref{block}. 
\FloatBarrier
\subsection{Calculating membership functions from cluster centers}
\label{sec:Calculating Mfs}
Only one CUDA kernel is defined to compute membership functions from the cluster centers. Rather than defining CUDA threads and block dimensions, the implementation in this kernel is quite similar to the sequential algorithm. The spawned CUDA threads are defined equally to the image pixels, which implies fine-grained granularity. Thus, one thread will handle one pixel. In correspondence to the previous phase of the proposed work \ref{block}, this kernel will be called in an iterative loop equally to the predefined number of clusters. At this stage, the computed new membership function arrays will be transferred to the host. The host will determine if the new membership function satisfies the condition as shown in Fig. \ref{block}. If the condition is satisfied, finally the cluster center arrays will be transferred back to the host. Defuzzification is performed and the the final segmented image is obtained. 
\FloatBarrier
\section{Implementation and Results}
\label{sec:results}
In this section, the implementation design of the proposed method is introduced in the first subsection. The functionality of the proposed method is proven using both qualitative and quantitative evaluations in the next subsections. The performance analysis is discussed in the final subsection.
\FloatBarrier
\subsection{Implementation}
\label{sec:implementation}
\begin{figure}[H]
\begin{center}
\includegraphics[scale=0.45]{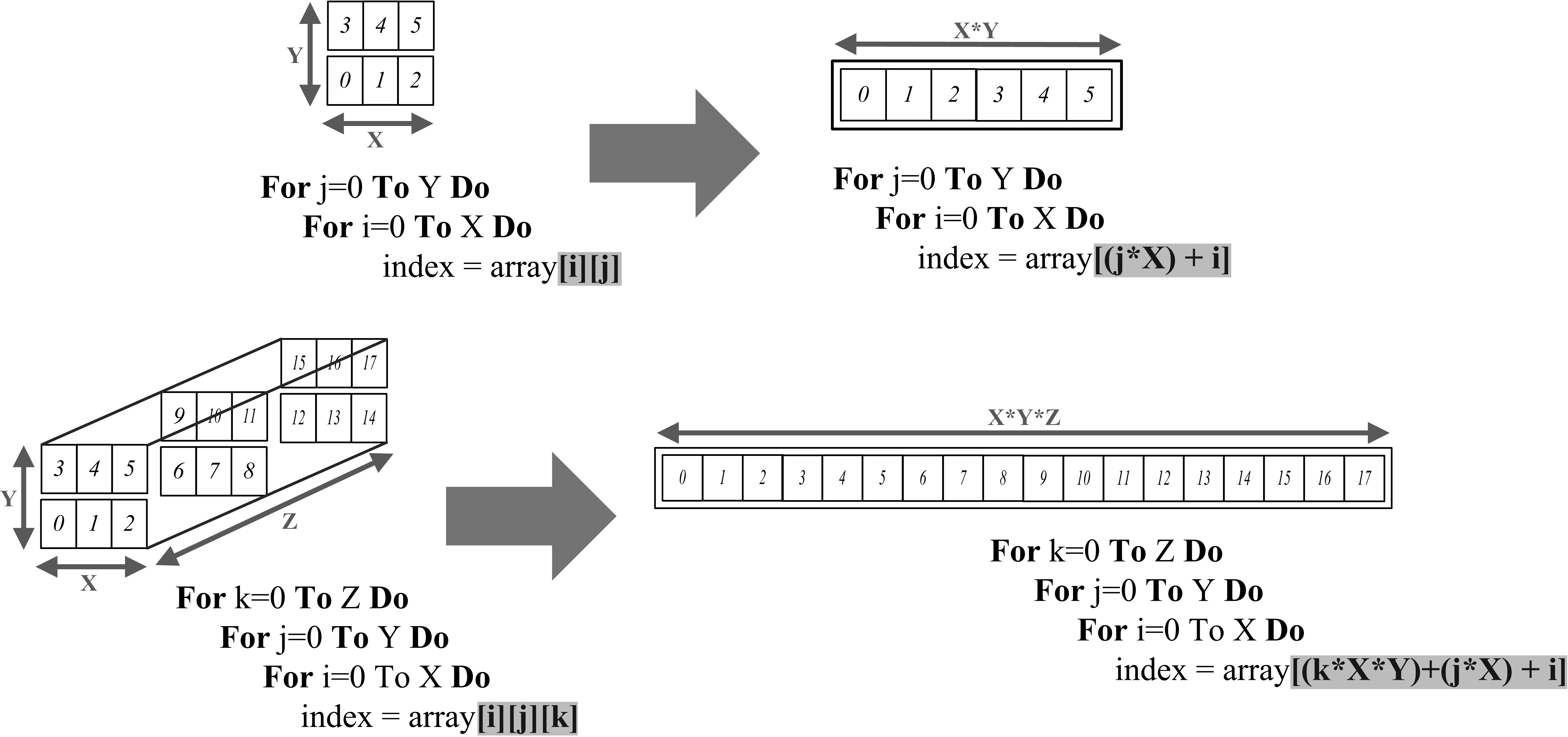}
\caption{Converting Multidimensional Arrays to One Dimensional}
\label{arrays}
\end{center}
\end{figure}

The proposed method was implemented using C language and CUDA. First, the sequential FCM algorithm was implemented in C. Our sequential C version was derived from a Java version available online at \cite{java}. The sequential FCM in C was tested on two different sequential platforms: Intel Core i5-480 CPU, Windows 7 Ultimate platform (we refer to as FCM1), and the second sequential platform is Intel Core 2 Duo CPU E7300, on Linux ubuntu 12.04 LTS (we refer to as FCM2). The reason of using two different sequential platforms in this work is to ensure the comparison is carried out between the fastest sequential FCM and the proposed parallel FCM.\\
\indent In the proposed parallel FCM, the image pixels, memberships, and cluster center arrays are defined in a 1-D pattern. The reason is to ensure coalesced memory transactions in the GPGPU. In addition, defining those input arrays in 1-D pattern will ease the number of CUDA block and grid sizes calculations. The CUDA block and grid sizes are consequently defined in 1-D patterns corresponding to the input arrays. Therefore, the form of the input has a significant effect on the performance of CUDA kernels because of the coalescing access \cite{Kirk2010}\cite{2014c}. Figure \ref{arrays} illustrates examples on the indices of arrays are modified when converting multi-dimensional arrays to 1-D arrays. In this work, the image array was converted from 2-D to 1-D, and the membership array was converted from 3-D to 1-D. The details of the parallel platform used in this experiment are shown in Table \ref{platform}.
\begin{table}[H]
\renewcommand{\arraystretch}{1.1}
\caption{Platform of the experiments} % title of Table
\centering  % used for centering table
\setlength{\tabcolsep}{5pt} % General space between cols (6pt standard)
\renewcommand{\arraystretch}{1} % General space between rows (1 standard)
\begin{tabular}{ l l } % centered columns (4 columns)
%\hline\hline                        %inserts double horizontal lines

\hline                  % inserts single horizontal line
 % inserting body of the table
    \textbf{CPU:}  & AMD Phenom(tm) II X4 810 Processor.  \\
    \textbf{Kernel:} & Linux  x86\_64 GNU.  \\
    \textbf{GPU:} & NVIDIA Tesla C2050.  \\
    \textbf{CUDA:} & CUDA compilation tools, release 5.0. \\
 % [1ex] adds vertical space
%\hline
\hline
\end{tabular}
\label{platform} % is used to refer this table in the text
\end{table}
\FloatBarrier
\subsection{Functionality Evaluation}
\label{sec:Functionality}
The proposed GPGPU-based FCM is tested on digital brain phantom simulated dataset from the Brain Web MR Simulator \cite{Collins} with the size of 20kB to segment white matter (WM), gray matter (GM) and cerebrospinal fluid (CSF) soft tissues regions. Skull stripping \cite{Dogdas} has been carried out on the brain phantom images to remove skull and other non-brain soft tissues, so that only brain soft tissues are used in the proposed parallel Fuzzy C-Means (FCM) segmentation process. When applying the proposed FCM on the brain soft tissues, four clusters are manually selected to represent the WM, GM, and CSF soft tissues regions and the final cluster represent the background region. Therefore in the proposed parallel FCM, there are four cluster center values being associated with the aforementioned regions. The functionality of the proposed method is then proven using both qualitative and quantitative evaluations in the following subsections. 
\FloatBarrier
\subsubsection{Qualitative evaluation}
\label{sec:Qualitative}
\begin{figure}[H]
\begin{center}
\includegraphics[scale=0.37]{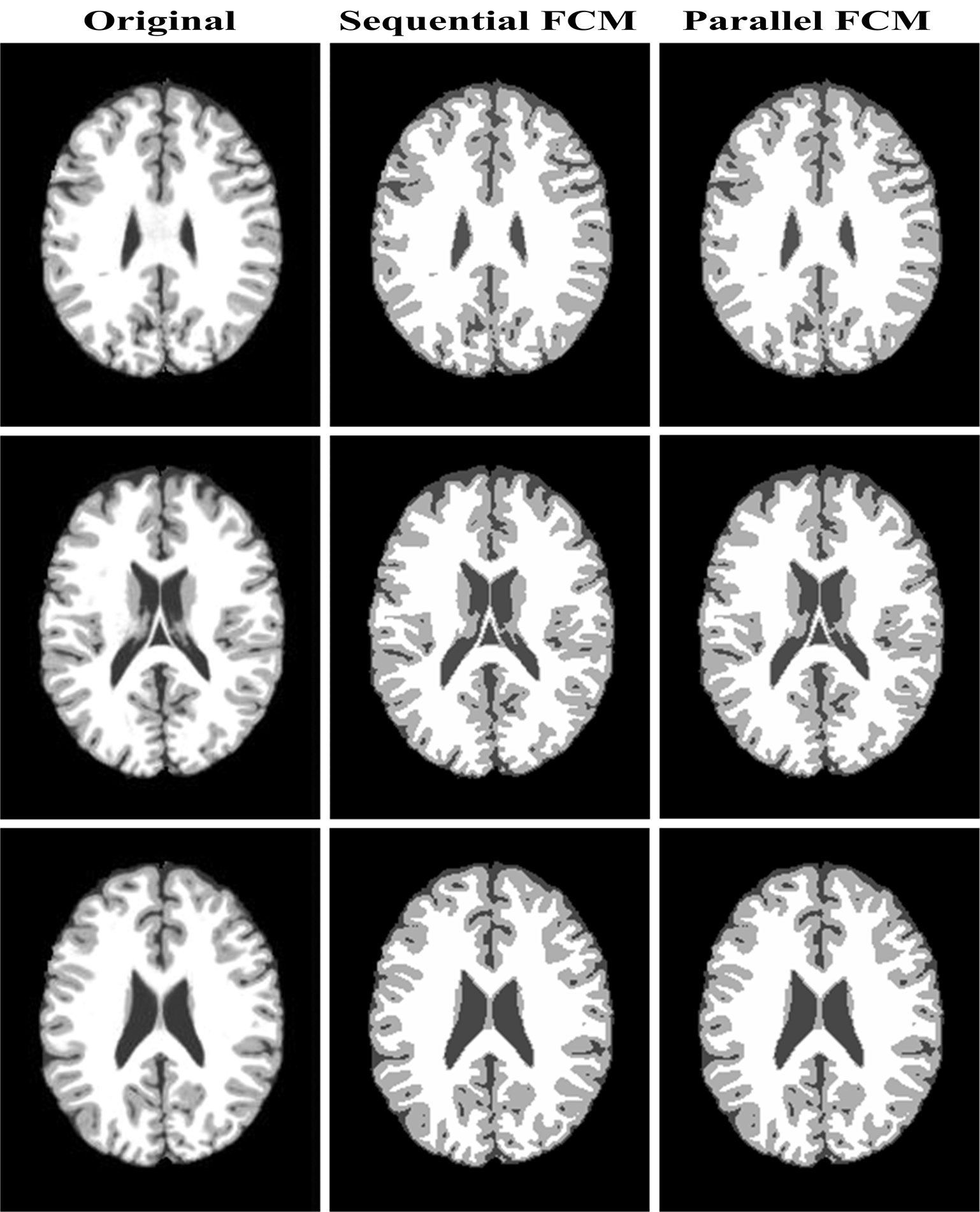}
\caption{Representative Results of The 101\textsuperscript{st}, 91\textsuperscript{st} and 96\textsuperscript{th} Axial Slice of Brain Tissue Phantom Using Sequential Fuzzy C-Means and The Proposed GPGPU-Based Fuzzy C-Means.}
\label{qual}
\end{center}
\end{figure}

The qualitative evaluation is performed for both the segmented results of the proposed parallel FCM and the sequential FCM. This is to evaluate the similarity of the segmented result of the proposed parallel FCM with the segmented result produced by the sequential FCM, visually. In Fig. \ref{qual}, the experiment results are presented. It can be seen that the result of the proposed method is identical to the result of the sequential FCM. 
%\cite{Database}
\FloatBarrier
\subsubsection{Quantitative evaluation}
\label{sec:Quantitative}
\begin{figure}[H]
\begin{center}
\begin{subfigure}{.15\textwidth}
\centering
  \includegraphics[width=.7\linewidth]{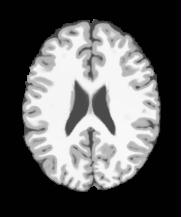}
  \caption{}
  \label{fig:sfig1}
\end{subfigure}\\
\begin{subfigure}{.15\textwidth}
\centering
  \includegraphics[width=.7\linewidth]{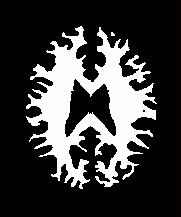}
  \caption{}
  \label{fig:sfig2}
\end{subfigure}
\begin{subfigure}{.15\textwidth}
\centering
  \includegraphics[width=.7\linewidth]{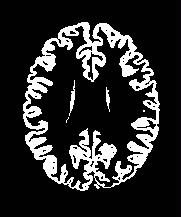}
  \caption{}
  \label{fig:sfig3}
\end{subfigure}
\begin{subfigure}{.15\textwidth}
\centering
  \includegraphics[width=.7\linewidth]{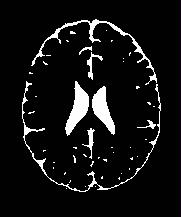}
  \caption{}
  \label{fig:sfig4}
\end{subfigure}
\begin{subfigure}{.15\textwidth}
\centering
  \includegraphics[width=.7\linewidth]{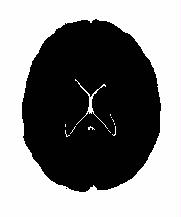}
  \caption{}
  \label{fig:sfig5}
\end{subfigure}
\caption{(a) The 96\textsuperscript{th} Axial Slice of Brain Tissues Phantom and The Corresponding Ground Truth Images (b) White Matter (c) Gray Matter (d) Cerebrospinal Fluid (e) Background.
}
\label{fig:fig}
\end{center}
\end{figure}

\begin{figure*}[t]
\begin{center}
\includegraphics[scale=0.31]{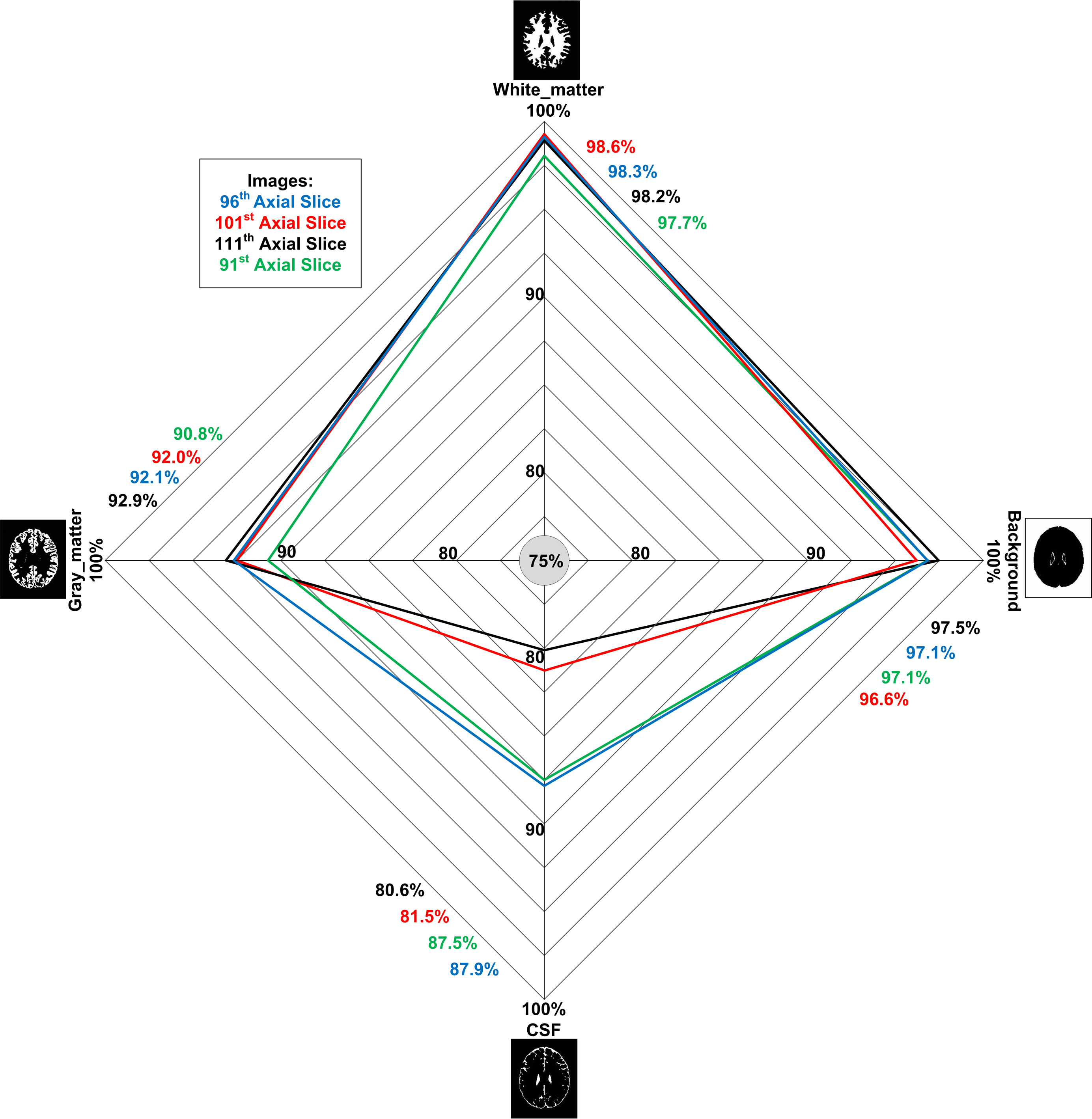}
\caption{Percentage of Dice Similarity Coefficient for 91\textsuperscript{th}, 96\textsuperscript{th}, 101\textsuperscript{th}, and 111\textsuperscript{th} Axial Slices of Brain Tissues Phantom Images}
\label{quan}
\end{center}
\end{figure*}

The quantitative evaluation is used to compare the results of the proposed parallel Fuzzy C- FCM and sequential FCM. Evaluation metrics such as Dice Coefficient Similarity (DSC) \cite{Zijdenbos} and performance analysis are used. DSC is used to evaluate if the accuracy of the segmented results of the proposed method is statistically similar to the segmented results of the sequential FCM based on the ground truth. While performance analysis is to compare the execution time and speed up of the proposed method with the sequential FCM. DSC is defined as in Equation \ref{eq116}.
\begin{equation}\label{eq116}
\large DSC=\frac{2(PR \cap GT)}{PR+GT}
\end{equation}
Where $PR$ is the segmented results of each method while $GT$ is the ground truth provided with the dataset \cite{Collins}. The DSC was implemented in C to be compatible with the implementation of the proposed method. An example of the ground truth is presented in Fig. \ref{fig:sfig5}.
Fig. \ref{quan} illustrates the percentage of DSC of the proposed parallel FCM and sequential FCM for white matter (WM), gray matter (GM), cerebrospinal fluid (CSF) and background regions for 91\textsuperscript{th}, 96\textsuperscript{th}, 101\textsuperscript{th} and  111\textsuperscript{th} axial slices of brain tissues phantom. The accuracy of the segmented results of both the proposed method and sequential FCM are statistically similar. 
\FloatBarrier
\subsection{Performance analysis }
\label{sec:Performance}
\begin{figure}[H]
\begin{center}
\includegraphics[scale=0.35]{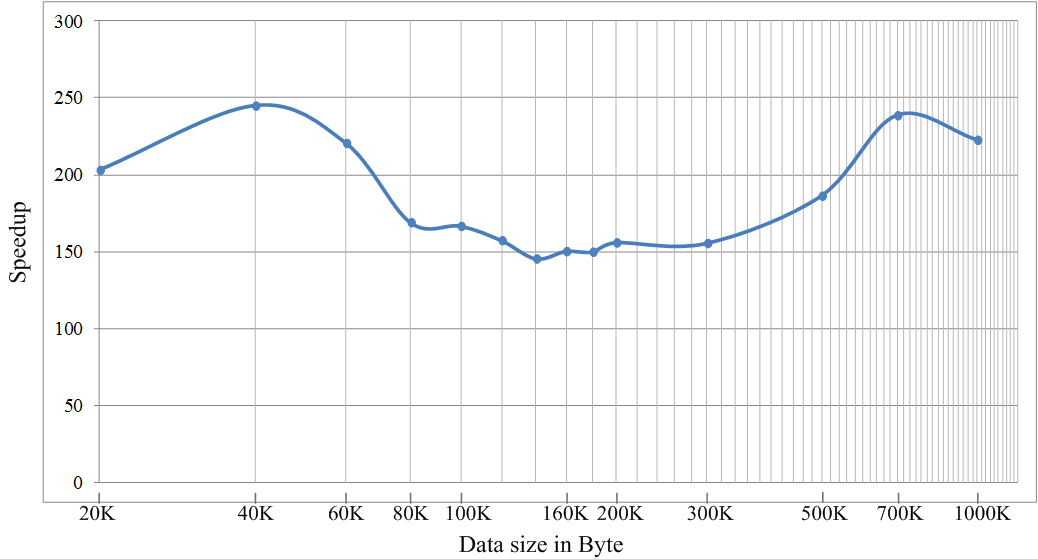}
\caption{Speed Up of The Proposed Parallel Fuzzy C-Means on Tesla C2050 of 448 Processing Elements}
\label{speedup}
\end{center}
\end{figure}

Once the functionality of the parallel approach is confirmed, performance analysis in terms of execution time and speedup was performed. As mentioned in Section \ref{sec:Calculating Mfs}, fine-grained granularity is adopted in this work in which one CUDA thread is spawned to manipulate one pixel. The total number of the spawned concurrent threads are equal to the image size to be segmented which indicates to the design scalability. The execution time was measured in both sequential and parallel approach for the process of calculating clusters centers and memberships. The initialization process was excluded from the measurements in both approaches. The function gettimeofday() was used to measure the elapsed time. For the sake of verification, the cudaEventRecord() function from CUDA API was also used to test the execution time. Table \ref{time} presents the execution time of the two sequential FCM (FCM1 and FCM2) and the proposed parallel FCM on GPGPU. The results of the execution time listed in Table \ref{time} and the corresponding speedup illustrated in Fig. \ref{speedup}, are the average execution time and speedup of 30 runs.\\
\indent From Table \ref{time}, it is shown that we have conducted experiments on various sizes of dataset from 20KB up to 1MB. In order to evaluate the execution time of the proposed parallel FCM in larger size dataset, we have enlarged the original phantom dataset 6KB (the original dataset size) up to 1MB. This enlargement is done only on the basis to evaluate the execution time of the proposed method in a larger size dataset. From Table \ref{time}, it is also shown that the execution time of the sequential FCM2 is slower compared to the sequential FCM1 for all the dataset. Therefore the results of FCM2 are compared over the results of the proposed parallel FCM. 
\begin{table}[H]
\renewcommand{\arraystretch}{1.1}
\caption{The Execution Time of The Sequential Fuzzy C-Means and The Proposed Parallel Fuzzy C-Means In Seconds.} % title of Table
\centering  % used for centering table
\setlength{\tabcolsep}{3pt} % General space between cols (6pt standard)
\renewcommand{\arraystretch}{1} % General space between rows (1 standard)
\begin{tabular}{ c  c  c  c } % centered columns (4 columns)
%\hline\hline                        %inserts double horizontal lines
\vtop{\hbox{\strut \textbf  {Dataset Size}}\hbox{\strut   ~~~~~(Byte)}}&\vtop{\hbox{\strut \textbf  {Sequential FCM1}}\hbox{\strut   ~~~~~~~~~(sec)}} &\vtop{\hbox{\strut \textbf  {Sequential FCM2}}\hbox{\strut   ~~~~~~~~~(sec)}} &\vtop{\hbox{\strut \textbf  {Parallel FCM}}\hbox{\strut   ~~~~~~~(sec)}}\\ [0.5ex] % inserts table 
%heading
\hline     % inserts single horizontal line
 % inserting body of the table
    20KB  & 57 & 10.16 &0.05  \\
    40kB & 114 & 19.6 & 0.08  \\
    60KB & 177 & 30.9 & 0.14  \\
    80KB   & 231 & 42.3 & 0.25  \\
    100KB   & 287 & 51.7 & 0.31  \\
    120KB   & 341 & 63 & 0.4  \\
    140KB   & 394 & 72.8 & 0.5  \\
    160KB   & 446 & 84.2 & 0.56  \\
    180KB   & 503 & 93 & 0.62  \\
    200KB   & 558 & 103 & 0.66  \\
    300KB   & 845 & 153 & 0.983  \\
    500KB   & 1420 & 261 & 1.4  \\
    700KB   & 1955 & 370 & 1.55  \\
    1000KB   & 2798 & 519 & 2.33 \\ [1ex]      % [1ex] adds vertical space
%\hline
%\hline
\end{tabular}
\label{time} % is used to refer this table in the text
\end{table}
\indent Fig. \ref{speedup} shows the speedup results of the proposed parallel FCM over the sequential FCM2. In Fig. \ref{speedup}, 245- to 169-fold speedup are obtained when the data size varies from 20 KB to 80 KB. When the data size is larger than 80 KB up to 300 KB, the speedup varies from 155- to 166-fold. 186- to 238-fold speedup recorded when the data size goes beyond 300 KB.\\
\FloatBarrier
\section{Conclusion and Future Works}
\label{sec:8}
GPGPUs are vary practical parallel models because they are affordable and not expensive. In this work, we proposed an efficient GPU-based implementation for Fuzzy C-Means algorithm. The functionality of the proposed parallel FCM has been verified and proven by conducting qualitative and quantitative evaluations. The empirical results show that the parallel FCM works precisely as the traditional sequential FCM. In addition, high performance and speedup of approximately 245 folds have been achieved compared with sequential FCM. \\
\indent Recently, new CUDA devices have been released featured with the capability of launching dynamic parallel kernels. Generally speaking, dynamic kernels or (nested kernels) enables to multiple levels reduction concurrently. It would be also an interesting topic in the future to implement FCM on such powerful devices.    
\FloatBarrier
\bibliographystyle{spbasic}      % basic style, author-year citations

\end{document}